\documentclass[useAMS,  usenatbib]{mn2e}
\usepackage{amsmath}
\usepackage{cases}
\usepackage{txfonts}
\usepackage{tipa}
\usepackage{amsfonts}
\usepackage{amssymb}
\usepackage{graphicx}

\title[Estimate Galactic Model Parameters]
{Estimation of Absolute Magnitude-dependent Galactic Model Parameters
In Intermediate Latitude With SDSS and SCUSS}

\author[Jia et al.]{
 Yunpeng Jia$^{1\star}$, Cuihua Du$^{1}$,
\thanks{E-mail:jiayunpeng11@mails.ucas.ac.cn(Jia); ducuihua@ucas.ac.cn(Du)}
Zhenyu Wu$^{2}$, Xiyan Peng$^{2}$, Jun Ma$^{2}$, Xu Zhou$^{2}$,
\and Xiaohui Fan$^{3}$, Zhou Fan$^{2}$, Yipeng Jing$^{4}$, Zhaoji Jiang$^{2}$, Michael Lesser$^{3}$, Jundan Nie$^{2}$,
\and Edward Olszewski $^{3}$, Shiyin Shen$^{5}$, Jiali Wang$^{2}$, Hu Zou$^{2}$, Tianmeng Zhang$^{2}$, Zhimin Zhou$^{2}$ \\
$^{1}$School of physics, University of the Chinese Academy of
Sciences, Beijing 100049, P. R. China\\
$^{2}$Key Laboratory of Optical Astronomy, National Astronomical Observatories, Chinese Academy of Sciences, Beijing, 100012, China\\
$^{3}$Department of Astronomy and Steward Observatory, University of Arizona, Tucson, Arizona, USA\\
$^{4}$Department of Physics and Astronomy, Shanghai Jiao Tong University, Shanghai 200240\\
$^{5}$Shanghai Astronomical Observatory, Chinese Academy of Sciences, Shanghai 200030 \\
}

\begin{document}

\date{Received}

\pagerange{\pageref{firstpage}--\pageref{lastpage}} \pubyear{2002}

\maketitle

\label{firstpage}

\begin{abstract}
Based on SDSS and South Galactic Cap of
u-band Sky Survey (SCUSS) early data, we use star counts method to
estimate the Galactic structure parameters in an intermediate
latitude with 10,180 main-sequence (MS) stars in absolute magnitude interval of $4 \leq M_r \leq 13$.
We divide the absolute magnitude into five intervals:$4 \leq M_r < 5$,
$5 \leq M_r < 6$, $6 \leq M_r < 8$, $8 \leq M_r < 10$, $10 \leq M_r \leq 13$, and
estimate the Galactic structure parameters in each absolute magnitude interval to
explore their possible variation with the absolute magnitude.
Our study shows the parameters depend on absolute magnitude.
For the thin disk, the intrinsic faint MS stars have large local space density
 and they tend to stay close to the Galactic plane.
A plausible explanation is that faint MS
stars with long lifetime experience long gravitational interaction time result in
a short scaleheight. However, for the thick disk, the parameters
 show a complex trend with absolute magnitude, which
 may imply the complicated original of the thick disk.
For the halo, the intrinsic faint MS stars have large local density and small
axial ratio, which indicate a flattened inner halo and a  more spherical
outer halo.
\end{abstract}


\section{INTRODUCTION}
The star counts method has been used to study the structure of
our galaxy by generations of astronomers. This useful method
can provide a measurement of the density distribution of the
stellar component of the Galaxy.  Since \citet{Bahcall1980}
fitted observations with two components Galactic model, namely disk and halo,
 and then improved by \citet{Gilmore1983} through introducing a third
component, namely the thick disk, the star counts method has
provided a picture of the standard Galactic model.
As the great improvement on data collection over the years,
more and more researchers try to refine both the estimation of standard Galactic model parameters
and the standard Galactic model to explain the currently available observations  well.
The better we know the Galactic structure, the better we understand the
formation and evolution of the Galaxy.

 A simple way to study Galactic structure is assuming a global smooth structure model of
our Galaxy, which is based on the modern of physics, and also on the pioneering work of
 \citet{Eggen1962}, who argued that the galaxy formed
from a relatively rapid ($\thicksim$$10^8$ years) radial collapse of the protogalactic cloud
 through studying the correlation between ultraviolet excess, metal abundance,
 angular momentum and the eccentricity of the Galactic orbit.
 The astronomers have done lots of works on the smooth structure parameters.
We can find elaborate list of the structural
parameters of the Milky Way in Table 1 of \citet{Chang2011}. We relist these
parameters in Table~\ref{table_ref} to conveniently find their improvements.
  Previous works show the estimation of the parameters
suffer from the degeneracy \citep{Robin00,Chen01,Siegel02,Phleps05,Juric08,Bilir08,Chang2011},
which could influence the confidence of the final results, but it is unconvinced to take
it as a main reason to explain the spread of values in Table~\ref{table_ref}.
We should notice the sample stars used in researchers' works are different. These parameters
 depend on both the properties and locations of the sample stars, such as
absolute magnitude \citep{Bilir06,Bilir2006,Karaali07},
Galactic longitude \citep{Du06,Cabrera-Lavers07,AK2007,
Bilir08,Yaz10,Chang2011} and Galactic latitude \citep{Du06, AK07},
thus these parameters that they derived could different from each other.
 The explanation of these dependence is still a topic of debate. The issue
 is not to give a explanation that could interpret the observation well,
  but rather to test whether
 the theory of the Galaxy is reasonable and therefore to answer a given question
 related to the Galactic formation and evolution.

Recent researches give us more information about the structure of the Galaxy:
the Galactic structure is not as smooth as we thought.
Many more substructures have been discovered, such as Sagittarius dwarf tidal stream
\citep[e.g.][]{Majewski2003}, Monoceros stream \citep[e.g.][]{Newberg2002}, and
Virgo overdensity \citep{Juric08}. The researches show
the Milky Way is a complex and dynamic structure that is still being shaped by the
merging of neighboring smaller galaxies, and the dynamic process may plays a crucial role
in establishing the galactic structure. These are reviewed in detail by
\citet{Ivezic12} and references therein. Needless to say,
the presence of irregular structure make the research about Galactic structure more complex.

We are interested in the global smooth structure of the Galaxy.
It can help us to define the
irregular structure, and then allow us to study the
formation and evolution of our Galaxy.
In this paper, based on SDSS and the South Galactic Cap u-band Sky Survey early data (SCUSS),
we attempt to study the structure parameters' possible variations with
absolute magnitude. We present density
function in Sect. 2. The introduction of SDSS and SCUSS and  the reduction of the data
would be described in Sect. 3.  Sect. 4 provide the method we used to
determine the parameters. Finally, our main results are discussed and
summarized in Sect. 5.

\begin{table*}
\begin{center}
\begin{tabular}{lllllllll}
\hline
$H_{z1}$ & $H_{r1}$ & $f_2$ & $H_{z2}$ & $H_{r2}$ & $f_h$ & Re(S) & $\kappa$ & Reference \\
(pc) & (kpc) & & (kpc) & (kpc) & & (kpc) & & \\ \hline
310-325  & -  &  0.0125-0.025    &   1.92-2.39 & -  &  -  & - &-  &   Yoshii (1982)  \\
300 & -  &   0.02    &   1.45    &   -  & -   &  -  &   -   &  Gilmore $\&$ Reid (1983) \\
325 & -    &   0.02    &   1.3     & -   &   0.002   &   3   &   0.85    &   Gilmore (1984) \\
280 &    -  &   0.0028  &   1.9 &    -  &   0.0012  &    -  &    -  & Tritton $\&$ Morton (1984) \\
125-475 &    -  &   0.016   &   1.18 - 2.21 &    -  &   0.0013  &    3.1*  &   0.8 & Robin $\&$ Creze (1986)   \\
300 &    -  &   0.02    &   1   &    -  &   0.001   &    -  &   0.85    &   del Rio $\&$ Fenkart (1987)  \\
285 &    -  &   0.015   &   1.3 - 1.5   &    -  &   0.002   &   2.36    &   Flat    &   Fenkart $\&$ Karaali (1987) \\
325 &    -  &   0.0224  &   0.95    &    -  &   0.001   &   2.9 &   0.9 &   Yoshii et al. (1987)  \\
249 &    -  &   0.041   &   1   &    -  &   0.002   &   3   &   0.85    &   Kuijken $\&$ Gilmore (1989) \\
350 &   3.8 &   0.019   &   0.9 &   3.8 &   0.0011  &   2.7 &   0.84    &   Yamagata $\&$ Yoshii (1992)    \\
290 &    -  &    -  &   0.86    &    -  &    -  &   4   &    -  & von Hippel $\&$ Bothun (1993)   \\
325 &    -  &   0.020-0.025  &   1.6-1.4 &    -  &   0.0015  &   2.67 &   0.8 &  Reid $\&$ Majewski (1993)    \\
325 &   3.2 &   0.019   &   0.98    &   4.3 &   0.0024  &   3.3 &   0.48    &   Larsen (1996)  \\
250-270 &   2.5 &   0.056   &   0.76    &   2.8 &   0.0015  &   2.44 - 2.75*    &   0.60 - 0.85 &  Robin et al. (1996, 2000)    \\
260 &   2.3 &   0.74   &   0.76    &   3 &    -  &    -  &    -  & Ojha et al. (1996)    \\
290 &   4   &   0.059   &   0.91    &   3   &   0.0005  &   2.69    &   0.84    &   Buser et al. (1998, 1999)    \\
240 &   - &   0.061   &   0.79    &   - &    -  &    -  &    -  & Ojha et al. (1999)    \\
280/267  &   -    &   0.02    &   1.26/1.29   &    -  &    -  &   2.99*   &   0.63    &  Phleps et al. (2000)  \\
330 &   2.25    &   0.065 - 0.13    &   0.58 - 0.75 &   3.5 &   0.0013  &    -  &   0.55    &   Chen et al. (2001)    \\
-   &   2.8 &   3.5     &   0.86        &   3.7     & -     & -         & -     & Ojha (2001) \\
280(350)    &   2 - 2.5 &   0.06 - 0.10 &   0.7 - 1.0 (0.9 - 1.2)   &   3 - 4  &   0.0015  &    -  &   0.50 - 0.70 &   Siegel et al. (2002)  \\
285 &   1.97    &    -  &    -  &    -  &    -  &    -  &    -  & Lopez-Corredoira et al. (2002)  \\
 -  &   3.5 &   0.02-0.03   &   0.9 &   4.7 &   0.002-0.003 &   4.3 &   0.5-0.6 &  Larsen $\&$ Humphreys (2003)  \\
320 &    -  &   0.07    &   0.64    &    -  &   0.00125 &    -  &   0.6 &   Du et al. (2003)  \\
265-495 &   -  &   0.052-0.098 &   0.805-0.970  &    -  &   0.0002-0.0015   &    -  &   0.6-0.8 & Karaali et al. (2004) \\
268 &   2.1 &   0.11    &   1.06    &   3.04    &    -  &    -  &    -  &   Cabrera-Lavers et al. (2005) \\
300  &   -    &   0.04-0.10    &   0.9   &    -  &    -  &   3/2.5*   &   1/0.6    &   Phleps et al. (2005)  \\
220 &   1.9 &    -  &    -  &    -  &    -  &    -  &    -  & Bilir et al. (2006b)  \\
160-360 &    -  &   0.033-0.076 &   0.84-0.87   &    -  &   0.0004-0.0006   &    -  &   0.06-0.08   &  Bilir et al. (2006a)  \\
301/259  &    -  &   0.087/0.055    &   0.58/0.93    &    -  &   0.001   &    -  &   0.74    &  Bilir et al. (2006c)   \\
220-320 &    -  &   0.01-0.07   &   0.6-1.1 &    -  &   0.00125 &    -  &   $>$0.4    &   \citet{Du06}  \\
206/198  &    -  &   0.16/0.10    &   0.49/0.58    &    -  &    -  &    -  &   0.45    &   Ak et al. (2007a)  \\
140-269 &    -  &   0.062-0.145   &   0.80-1.16   &    -  &    -  &    -  &    -  &  Cabrera-Lavers et al. (2007)   \\
220-360 &   1.65-2.52   &   0.027-0.099   &   0.62-1.03   &   2.3-4.0 &   0.0001-0.0022   &    -  &   0.25-0.85   & Karaali et al. (2007) \\
167-200 &    -  &   0.055-0.151 &   0.55-0.72   &    -  &   0.0007-0.0019    &    -  &   0.53-0.76   &  Bilir et al. (2008)    \\
245(300)    &   2.15(2.6)   &   0.13(0.12)  &   0.743(0.900)    &   3.261(3.600)    &   0.0051  &   2.77*   &   0.64    &   \citet{Juric08}   \\
325-369 &   1.00-1.68   &   0.0640-0.0659   &   0.860-0.952 &   2.65-5.49   &   0.0033-0.0039   &    -  &   0.489-0.654   &  Yaz $\&$ Karaali (2010)  \\
360 &   3.7 &   0.07    &   1.02    &   5   &   0.002   &   2.6*    &   0.55    &   Chang et al. (2011)   \\
103-350 & - & 0.083-0.165 & 0.525-0.675 & - & 0.0005-0.0065 & - & 0.20-0.84 & this work \\ \hline
\end{tabular}
\end{center}
\caption{Galactic model parameters tabulated base on
the Table 1 in \citet{Chang2011}. $H_z$ and $H_r$ mean scaleheight and
scalelengh, respectively. And suffix 1 and 2 denote thin disk and thick disk,
repectively. $f_2$ and $f_h$ are the local stellar density ratio of
the thick-to-thin disk and halo-to-thin disk, respectively.
 The parentheses are the corrected values for binarism.
The asterisk denotes the power-law index replacing Re, which is commonly
known as the de Vaucouleurs radius, and $\kappa$ is axial ratio.
} \label{table_ref}
\end{table*}

\section{DENSITY LAWS}

In this study, we adopt the density laws of disk  by two exponentials functions
in cylindrical coordinates:
\begin{equation} \label{disk_law}
  D_i(x,z)=n_i\, exp(-(x-R_\circleddot)/l_i)\, exp(-(|z|-z_\circleddot)/h_i)
\end{equation}
where $z=z_\circleddot + rsin\, b$ is the vertical distance
to the Galactic plane, $b$ is Galactic latitude, $r$ is photometric distance,
$z_\circleddot$ is the vertical distance of the Sun to the
plane (25 $pc$, Juri\'c et al. 2008). $x$ is
the distance to Galactic center on the plane.
$R_\circleddot$ is the distance of the Sun to Galactic center (8 kpc, Reid 1993).
$l_i$ and $h_i$ are the saclelength and scaleheight, respectively,
and $n_i$ is the normalized number density at $(R_\circleddot,z_\circleddot)$,
suffix $i$ take the values 1 and 2 for thin disk and thick disk, respectively.

The density law for halo used most is the de Vaucouleurs spheroid (1948) which was used to
describe the surface brightness profile of elliptical galaxies, known as $r^{1/4}$ law.
According to \citet{Young1976}, the projection of this law into three dimensions
has no simple analytic form, but it can derive the asymptotic expansion at origin
and infinity point. Based on it, \citet{Bahcall1986} gave an
analytic approximation:
\begin{align} \label{eq2}
D_3(R,b,l) =n_3 \, (R/R_\circleddot)^{-7/8}\, \{exp[-10.093(R/R_\circleddot)^{1/4}+10.093]\}
\nonumber \\
 \times [1-0.08669/(R/R_\circleddot)^{1/4}], \, R \geq 0.03R_\circleddot
\nonumber \\
 \times 1.25(R/R_\circleddot)^{-6/8} \, \{exp[-10.093(R/R_\circleddot)^{1/4}+10.093]\},
\, R < 0.03R_\circleddot
\end{align}
where $R$ is Galactocentric distance:
\begin{flushleft}
 $R=(x^2+(z/\kappa)^2)^{1/2}$,\\
$x=[R_\circleddot^2 +R^2cos^2b-2R_\circleddot R cosb cosl]^{1/2}$,
\end{flushleft}
and $\kappa$ is the axial ratio,
$b$ and $l$ are the Galactic latitude and longitude.

\section{DATA REDUCTIONS}

\subsection{SCUSS and SDSS}

The South Galactic Cap u-band Sky Survey (SCUSS) is an international
cooperative project between National Astronomical Observatories,
 Chinese Academy of Sciences and Steward Observatory, University of Arizona, USA.
 The project plan to perform a sky survey of about 3700 square degree
 field of the south Galactic cap in u band (3508 \AA) with
 the 90 inch (2.3 meter)  Bok telescope. And this project will also provide part of
the essential input data to the The Large Sky Area Multi-Object Fiber
Spectroscopic Telescope (LAMOST) project.
The telescope equipped with
 four $4K \times 4K$ CCD mosaic (64-megapixel) and the field of view
 is 1.08 $deg$ $\times$ 1.03 $deg$. The exposure time is 5 minutes
 and the limiting magnitude will reach about 23 mag (signal-to-noise = 5).
 The u band filter, which been used in SCUSS project, is almost same
 as SDSS u band filter but has few blue shift
 and do not need any color items of magnitude conversion between
 standard SDSS u band and SCUSS u band photometric
 system at first. By testing  the data
obtained from previous camera and new 90 Prime camera of BOK telescope,
the limiting magnitude of SCUSS u band may be 1.5 mag deeper than
SDSS u band and can be calibrated by SDSS u band in about 2/3 of
 SCUSS field. The SCUSS can be used to study Star formation rate,
 Galactic interstellar extinction,
 Galaxy morphology, the Galaxy structure, Quasi-Stellar Object, Variable star and Cosmology.
 Here we just use the early data to study the Galactic structure,
 and the SCUSS data is still updating.
The detailed survey overview and data reduction about SCUSS are in
preparing (Zhou et al. 2014; Zou et al. 2014).

The Sloan Digital Sky Survey (SDSS) is a large international collaboration project,
it has obtained deep, multi-color images covering more than a quarter of the sky.
The SDSS used a dedicated 2.5-meter telescope at Apache Point Observatory, New Mexico.
equipped with two instruments: a CCD camera with 30 2048 $\times$ 2048 CCDs in five
filters \citep[$ugriz$;][]{Fukugita1996} and two 320 fiber double spectrographs.
The flux densities are measured in five bands ($u$, $g$, $r$, $i$, $z$) with
effective wavelengths of 3551, 4686, 6165, 7481
and 8931 \AA, respectively. The $95\%$ completeness limits of the images are
u, g, r, i, z = 22.0, 22.2, 22.2, 21.3, 20.5, respectively \citep{Abazajian2004}.
The imaging data are automatically processed through a series of
 software pipelines to produce a catalogue.
 The relative photometric calibration accuracy for $u$, $g$, $r$,
 $i$, $z$ are $2\%$, $1\%$, $1\%$, $1\%$ and $1\%$, respectively \citep{Padmanabhan08}.
 The detailed information about SDSS can be found on the SDSS web site
(\emph{http://www.sdss.org}) or one can see the overview of SDSS in
  e.g. \citet{Stoughton2002}, \citet{Abazajian2009} and \citet{Aihara2011}.

In Table~\ref{tab_compare}, we list the parameters of SCUSS and SDSS
filters. Column (1) represents the ID of SCUSS and SDSS filters, and
columns (2) and (3) represent effective wavelengths and full width at
half-maximum (FWHM) of six filters, respectively.

\begin{table}
\begin{center}
\begin{tabular}{p{1.8cm}p{1.8cm}p{1.8cm}}
\hline
\hline
Filter & Wavelength & FWHM \\
       & (\AA)    &(\AA)\\ \hline
 u (SCUSS)  & 3508   &  360      \\ \hline
 u (SDSS)   & 3551   &  570  \\ \hline
 g   & 4686   &  1390   \\ \hline
 r   & 6165   &  1370  \\ \hline
 i   & 7481   &  1530  \\ \hline
 z   & 8931   &  950   \\ \hline
\hline
\end{tabular}
\caption{Parameters of SCUSS and SDSS filters.
Column (1) represents the ID of SCUSS and SDSS filters, and
columns (2) and (3) represent effective wavelengths and full width at
half-maximum (FWHM) of six filters, respectively.
} \label{tab_compare}
\end{center}
\end{table}

\subsection{Choose sample stars}
The data used in this work are taken from SCUSS early data and
the eighth data release of SDSS \citep{Aihara2011}. We choose 7.07 $deg^2$
field locate at $50 ^\circ \leq l \leq 55^\circ$, $-46 ^\circ \leq b\leq -44^\circ $.
Considering the SDSS CCDs saturated at about
$r \sim 14$ mag \citep{Ivezic01} and the robust point source-galaxy separation at
$r \sim 21.5$ mag \citep{Lupton02}, we  restrict the apparent magnitudes in the range of
$15 \leq r \leq 21.5$. As SCUSS can reach deeper magnitude on u-band than SDSS,
we use u band data from SCUSS early observation, including apparent magnitude
 u and the corresponding errors, instead of those from SDSS.  The
u band magnitude and error mentioned hereafter will be the ones from SCUSS. The total
absorptions for each star, $A_u$, $A_g$, $A_r$, $A_i$, $A_z$, are  taken
from SDSS. The apparent magnitudes $u$, $g$, $r$, $i$, $z$ are
de-reddened:
\begin{displaymath}
u_0=u-A_u, \; g_0=g-A_g, \; r_0=r-A_r, \;
i_0=i-A_i, \; z_0=z-A_z
\end{displaymath}
All the magnitudes and colors mentioned hereafter will be de-reddened ones.

Star/Galaxy classification base on the `type' parameter provided by SDSS
(the value 3 means Galaxy, and 6 means Star), the technical details can be
found on the SDSS web site (\emph{http://www.sdss.org}).
\textbf{In Figure 1,  we give the two-color diagram  $(r-i)_0$ versus $(g-r)_0$ distribution of all stars
and the solid line is stellar locus which is described by Eq.~\ref{eq3} (Juri\'c et al. 2008),
the dashed line and the dotted-dashed line are 0.15 mag and 0.3 mag from stellar locus,
respectively. As shown in Fig.~1, there are some non-main sequence (non-MS) stars.
We apply the method mentioned in \citet{Juric08}
to remove hot white dwarfs, low-redshift quasars and white dwarf/red dwarf
unresolved binaries from our sample.
Their procedure consists of rejecting objects at distances larger than 0.3 mag
from the stellar locus.
But as pointed by \citet{Yaz10}, rejecting objects at distance larger than 0.15 mag
is appropriate for Galactic latitude $<b>=45^\circ$.}
In this study, we reject those objects at distance larger than 0.15 mag.
In addition, we also restrict the objects in interval $(g-r)_0 \leq -0.3$
to remove very early MS stars (see Fig.~2 in Yaz \& Karaali 2010) and
in interval  $(r-i)_0 \geq 1.8$ to ensure the photometric parallax method
is good enough (see Figure 2 in Juri\'c et al. 2008).

\begin{align} \label{eq3}
(g-r)_0 =& 1.39\{1-exp[-4.9(r-i)_{0}^3-2.45(r-i)_{0}^2 \nonumber \\
 & -1.68(r-i)_{0}-0.05]  \}
\end{align}
While the two-color diagram of $(u-g)_0$ versus $(g-r)_0$ in Fig.~\ref{figugr}(a),
show there still exist non-MS stars. We use Eq.~\ref{eq4} to roughly remove those:
\begin{equation} \label{eq4}
(u-g)_{0}=exp[-(g-r)_{0}^2+2.8(g-r)_{0}-1]
\end{equation}
Any point source lie at the distance in vertical direction to the locus that described
by Eq.~\ref{eq4} further than 0.6 mag would be rejected. Simultaneously, we reject the point source
whose $(u-g)_0$ is smaller than 0.6 mag to roughly remove quasars
 (see Fig.~9 in Juri\'c et al. 2008 and Richards et al. 2002).

The absolute magnitude can be derived by photometric parallax method
described by \citet{Juric08}, i.e. the ``bright normalization'' :
\begin{align} \label{eq5}
M_r =& 3.2+13.30(r-i)_{0}-11.50(r-i)_{0}^2  \nonumber \\
 & +5.4(r-i)_{0}^3-0.70(r-i)_{0}^4
\end{align}
Note that this relation was even constructed with
kinematic consideration to reconcile the
differences between some relations proposed by different researchers
(see Fig.~2 in this paper). We must point that the effect
of the metallicity on this photometric parallax relation is hard to correct
satisfactorily (Juri\'c et al. 2008 have illuminated the reason in detail),
especially the gradients of the metallicity distribution
are exist in Galactic components
\citep[e.g.][]{Ivezic2006,Bilir2012,Coskunoglu2012,Peng2012, Peng2013}.
After calculating the absolute magnitude, the photometric distance $r$
 is derived from the following equation:
\begin{equation}
r_0-M_r=5logr-5
\end{equation}

In Figure~\ref{fighist}, we show the apparent magnitude (r-band)
histogram of sample stars.
 Two arrows show the limited r-magnitude in our work:
15.0 $mag$ for lower limit and 21.5 $mag$ for upper limit.
The grey area represents the distribution of our final sample stars in
absolute magnitude range from 4 $mag$ to 13 $mag$.

As a summary, the data reduction process should be as precisely as possible to
remove non-MS points, but either  over reduced or  less reduced
is exists in practical work.
However, as long as the non-MS points are far less than MS stars,
we consider the overall distribution of sample stars could smooth the influence caused
by non-MS stars,
thus it would contribute little to the effect on the determination of parameters.
\begin{figure}
\includegraphics[angle=-90,width=0.45\textwidth]{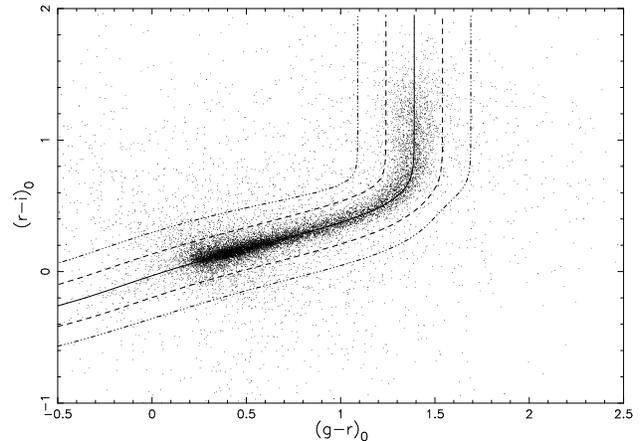}
\vspace{+0.7cm}
\caption{The distribution of stars in $(g-r)_0$ versus $(r-i)_0$.
The solid line is stellar locus,
the dot-dashed lines means 0.3 mag from stellar locus, and dashed lines
means 0.15 mag from stellar locus. In this study, we  reject
objects further than 0.15 mag from the stellar locus. } \label{figgri}
\end{figure}

\begin{figure}
\includegraphics[angle=-90,width=0.45\textwidth]{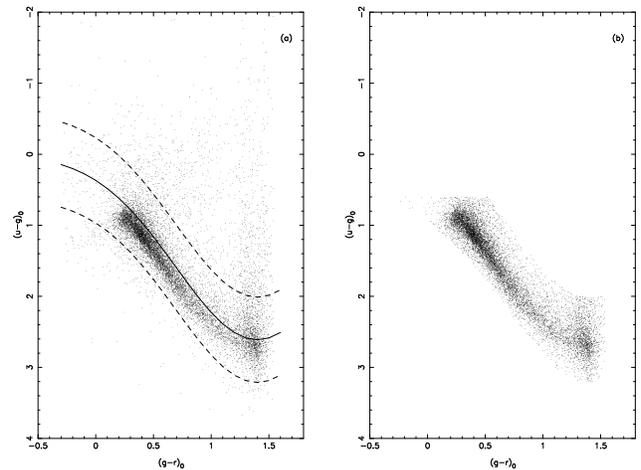}
\vspace{+0.7cm}
\caption{The distribution of stars in $(u-g)_0$ versus $(g-r)_0$ after
removing some non-MS stars in  $(g-r)_0$ versus $(r-i)_0$. In Figure (a),
the solid line described by Eq.~\ref{eq4}, and the dashed lines describe
the points whose distance in vertical direction to the solid line
 are 0.6 mag. Figure (b) describe the final distribution of stars
 in  $(u-g)_0$ versus $(g-r)_0$. } \label{figugr}
\end{figure}

\begin{figure}
\includegraphics[angle=-90,width=0.45\textwidth]{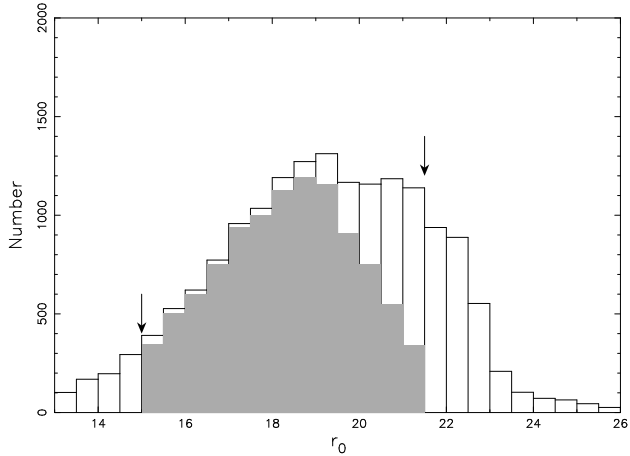}
\vspace{+0.7cm}
\caption{The apparent magnitude histogram of sample stars in r-magnitude. Two arrows
show the bright and faint limiting r-magnitudes in our work. The grey area is our final
sample stars which distributed in absolute magnitude from 4 $mag$ to 13 $mag$.
 } \label{fighist}
\end{figure}

\section{GALACTIC MODEL PARAMETERS }

In this study, we choose MS stars in absolute magnitude interval
$4 \leq M_r \leq 13$ to estimate the parameters by
minimising the reduced $\chi^2$. All the parameters errors estimations are obtained by
increasing $\chi^2_{min}$ by 1 \citep{Phleps00}.

\subsection{The procedure used in this work}

 (\romannumeral 1) We divide the absolute magnitude into five intervals:
 $4 \leq M_r < 5$, $5 \leq M_r < 6$, $6 \leq M_r < 8$,
$8 \leq M_r < 10$, $10 \leq M_r \leq 13$, see Table~\ref{tab1}.
This way of dividing the intervals can ensure enough numbers of sample stars in each interval
to explore possible parameters variations with absolute magnitude.
(\romannumeral 2)
We divide the distance into suitable numbers of logarithmic bins,
and then derive the observed number density distribution in each absolute magnitude
 interval by using the following equations:
\begin{eqnarray}
D(r) & = & \frac{N(r)}{\Delta V_{12}} \\
\Delta V_{12} & = & (\frac{\pi}{180})^2 \frac{\omega}{3} (r^3_2-r^3_1) \\
\sigma_D(r) & = & \frac{\sqrt{N(r)}}{\Delta V_{12}}
\end{eqnarray}
where $\omega$ denotes the size of the field (unit in $deg^2$ ), $r_1$ and
 $r_2$ are lower limit and upper limit of the volume $\Delta V_{12}$, respectively,
and N is the number of stars, $\sigma_D$ is Poisson error of the density
due to the shot noise.
For every logarithmic distance bin we use the mean height $z$ above the Galactic plane
$z_i = \sin b <r_i>$, where $<r_i>=[(r_i^3 + r_{i+1}^3)/2]^{1/3}$.
(\romannumeral 3)
We obtain the parameters by fitting the model to the observation with $\chi^2$ method.
In the last process, we distinguish disk and halo by their
spatial distribution if the sample stars in the considered absolute
magnitude interval are composed of three components (see Table~\ref{tab1}). In this
condition, based on the pioneering works \citep[e.g.][]{Reid-Majewski1993,Bilir06},
we take $z=4$ kpc  as the boundary of disk and halo, and
simultaneously derive the parameters of thin and thick disk first,
then fit thick disk and halo with the overall density distribution
using the determined value of thin disk parameters to obtain the parameters of
the last two components as  final results. Note that the thick disk parameters are
derived twice. In order to distinguish between the two ways in obtaining the thick disk
parameters, we call the way fitting with thin and thick disks as `disk-fitting', and
the way fitting with thick disk and halo as `halo-fitting'.

\begin{table}
\begin{tabular}{ccccc}
\hline
Absolute magnitude & Number & $r_{min}$ & $r_{max}$ & component \\ \hline
$4 \leq M_r < 5$ & 2680 & 1000 pc & 31.6 kpc & thick+halo \\ \hline
$5 \leq M_r < 6$ & 3213 & 631 pc  & 19.9 kpc & thin+thick+halo \\ \hline
$6 \leq M_r < 8$ & 2568 & 151 pc  & 12.6 kpc & thin+thick+halo \\ \hline
$8 \leq M_r < 10$ & 1258 & 100 pc & 5.0 kpc & thin \\ \hline
$10 \leq M_r \leq 13$ & 461 & 25 pc & 2.0 kpc & thin \\ \hline
\end{tabular}
\caption{Informations about sample stars in each absolute magnitude interval.
The second column denotes the number of sample stars. The third and fourth column
are the lower and upper limit distance, repectively. The last column means
the components we considered in each absolute magnitude interval.
 Even though the distance can up to 5.0 kpc in interval
$[8-10)$, the thick disk stars are infrequent, we just consider the
thin disk only. The similar condition hold for the thin disk in interval $[4-5)$.
} \label{tab1}
\end{table}

\subsection{Absolute magnitude-dependent Galactic model parameters}

As the scalelength is in magnitude of kiloparsec, compare with scaleheight in parsec
 (see Table~\ref{table_ref}),
 which is so large that contribute
little to the change of the density in a small field, as shown in
Eq.~\ref{disk_law}, it has little impact on other parameters, so
 we roughly evaluate them in absolute magnitude
interval $4 \leq M_r \leq 13$ by disk-fitting, and hereafter fix them when obtaining
the other parameters in every interval.

Our results shows the values of $l_1$ and $l_2$ are in the range of $l_1$=2400--4800 pc,
and $l_2$=3600--6000 pc, respectively. The poor constraint on scalelengh probably due to
our sample stars distributed in a narrow radial direction of cylindrical coordinate.
As other parameters are insensitive to scalelengh for the reason cited above,
 we take $l_1$=3000 pc and $l_2$=5000 pc in this work, which are roughly equal with
 the median value of the range they spread.

With the evaluated value of scalelength, we quickly obtain other parameters
in each absolute magnitude interval. In order to compare with other researchers' works,
we also obtain the parameters in interval $4 \leq M_r \leq 13$.
Our final results of parameters are listed in Table~\ref{tab2}.

\begin{figure*}
\includegraphics[angle=-90,width=0.8\textwidth]{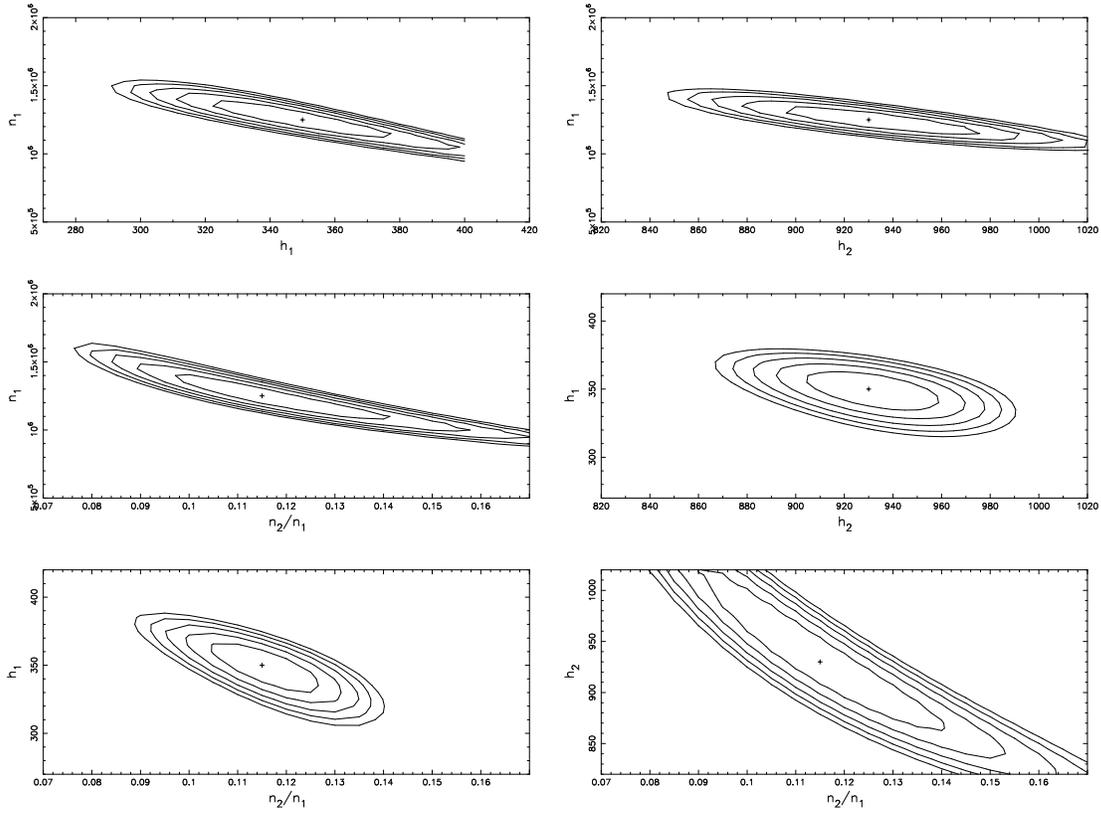}
\vspace{+0.7cm}
\caption{Contour plotted of the $\chi^2$ value of pairs of disk (thin and
thick disk) parameters
in absolute magnitude interval $5 \leq M_r < 6$.
The cross mark represents the best-fit values of the Galactic model parameters.
 } \label{fig_diskcontour}
\end{figure*}

For the thin disk,  with the increase of absolute magnitude, the scaleheight
 decreases from 350 $pc$ to 103 $pc$, while the local density increases from
$1.25\times10^6$ $star/kpc^3$ to $2.32\times10^7$ $star/kpc^3$. The above results are consistent
with \citet{Bilir06} (see Table 6 in that paper). The results reveal a
phenomenon for thin disk: the intrinsic faint MS stars tend to stay closer
to Galactic plane and have larger local density than that of intrinsic bright MS stars.
Consequently, as the contribution of the faint stars in our work,  the scaleheight
of the thin disk ($h_1 =$205 $pc$) in interval $4 \leq M_r \leq 13$ is smaller
than most previous works. As shown in Figure~\ref{fig_diskcontour},
the dependence between the disk parameters is exist in this work.
 This figure is plotted by disk-fitting in interval $5 \leq M_r < 6$,
other intervals also plot similar shape.  The cross mark represents
the best-fit values of the Galactic model parameters.

\begin{figure*}
\includegraphics[angle=-90,width=0.8\textwidth]{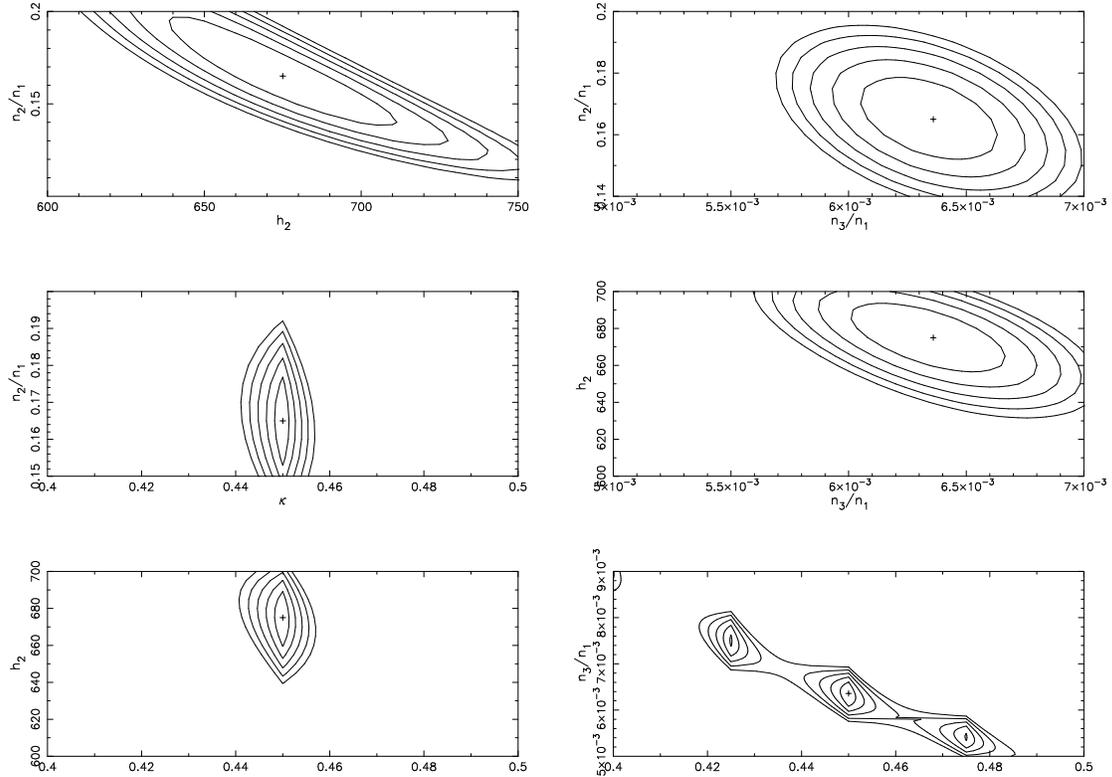}
\vspace{+0.7cm}
\caption{Contour plotted of the $\chi^2$ value of pairs of thick disk and
halo parameters
in absolute magnitude interval $5 \leq M_r < 6$.
The cross mark represents the best-fit values of the Galactic model parameters.
 } \label{fig_halocontour}
\end{figure*}

However, from Table~\ref{tab2} we notice that the thick disk parameters are non-monotonic
change with absolute magnitude . We also notice the dependence between
$n_2$ and $h_2$: the larger $n_2$, the smaller $h_2$. The same dependence also holds
between $n_1$ and $h_1$. This is
 caused by the intrinsic property of Eq.~\ref{disk_law}:
 $h_2$ and $n_2$, $n_1$ and $h_1$ are anti-correlation.
 These dependence will emerge when fitting the model to observation, which
  are also showed in Figure~\ref{fig_diskcontour}. In addition,
we can see the dependence between thick disk parameters
in Figure~\ref{fig_halocontour}, which is plotted by the halo-fitting.
For the disk-fitting,
 the scaleheight $h_2$ and local density normalized to thin disk $n_2/n_1$
is 900 $pc$ and 0.115, respectively, in absolute magnitude interval
$5 \leq M_r < 6$. While for the halo-fitting, $h_2$ and ${n_2}/n_1$ is
675 $pc$ and 0.165, respectively.
The comparison of the two sets of parameters
is listed in Table~\ref{tab3}. Notice that the scaleheight of halo-fitting is always
smaller than disk-fitting, this is caused by the contribution from the halo stars
at large distance in the disk-fitting, which can overestimate the scaleheight.
 We can also notice that
the large discrepancy exists between the two pairs of
thick disk parameters in interval $5 \leq M_r < 6$, a possible reason  is
 the degeneracy exists in parameters of halo (see below),
 which could influence the determination of thick disk parameters
(see Figure~\ref{fig_halocontour}).
 As the halo-fitting consider the contribution of the halo, thus
the parameters obtained in this way are more reliable,  so we
favor the value of thick disk parameters  obtained by halo-fitting than
it did by disk-fitting.

For the halo, with the increase of absolute magnitude,
 the local density $n_3$ increases from $1.54\times10^3$ $star/kpc^3$ to
$3.01\times10^4$ $star/kpc^3$, while the axial ratio $\kappa$ decreases from 0.84 to 0.20
(see Table~\ref{tab2}).
The large range of $\kappa$ is also found by \citet{Karaali07},
who gave the range of $0.25 \leq \kappa \leq 0.85$, which is
consistent with this work. The results show intrinsic fainter MS stars in halo
have larger local density and their distributions are more non-spherical.
As  large distance can be reached in bright interval,  so the
values of $\kappa$ shows the halo is more spherical at large distance,
indicates a flattened inner halo and a spherical outer halo. The value of
$\kappa$ in interval $4 \leq M_r \leq 13$ is 0.55, which is closed
to the recent results \citep{Du06,Bilir06,Juric08,Chang2011}. The contour
plot of halo-fitting is shown in Figure~\ref{fig_halocontour}.
In the bottom right of this Figure, $\kappa$ versus $n_3/n_1$ shows
the degeneracy exist between this two parameters. The similar degeneracy
between this two parameters is also found in interval $6 \leq M_r < 8$
and $4 \leq M_r \leq 13$, but
not found in interval $4 \leq M_r < 5$. Needless to say, the degeneracy
could influence the parameters of the thick disk
and halo in this work.

\begin{figure*}
  \centering
  \begin{minipage}[b]{0.45\linewidth}
    \centering
    \includegraphics[angle=-90,width=0.8\textwidth]{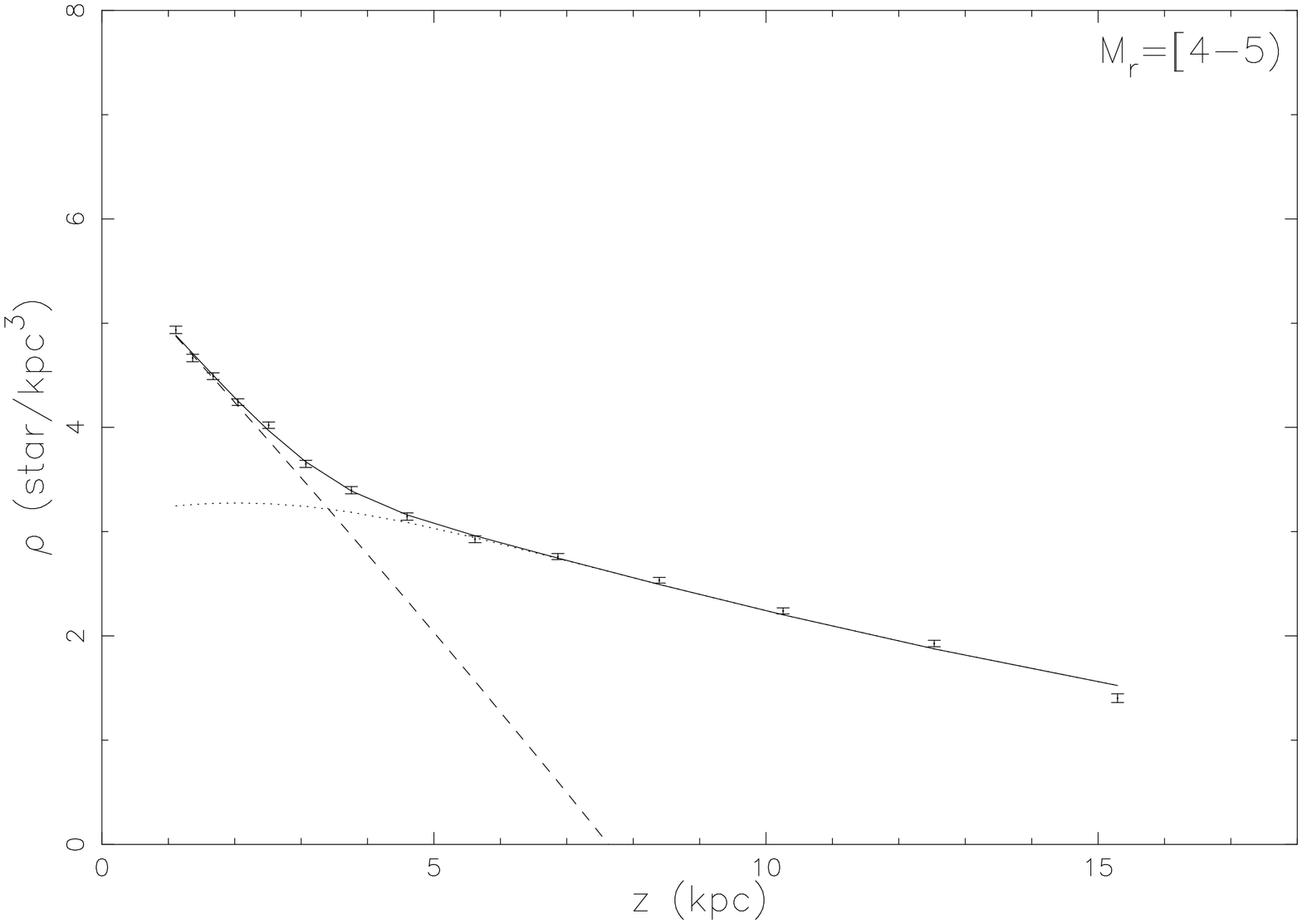}
  \end{minipage}%
  \begin{minipage}[b]{0.45\linewidth}
    \centering
    \includegraphics[angle=-90,width=0.8\textwidth]{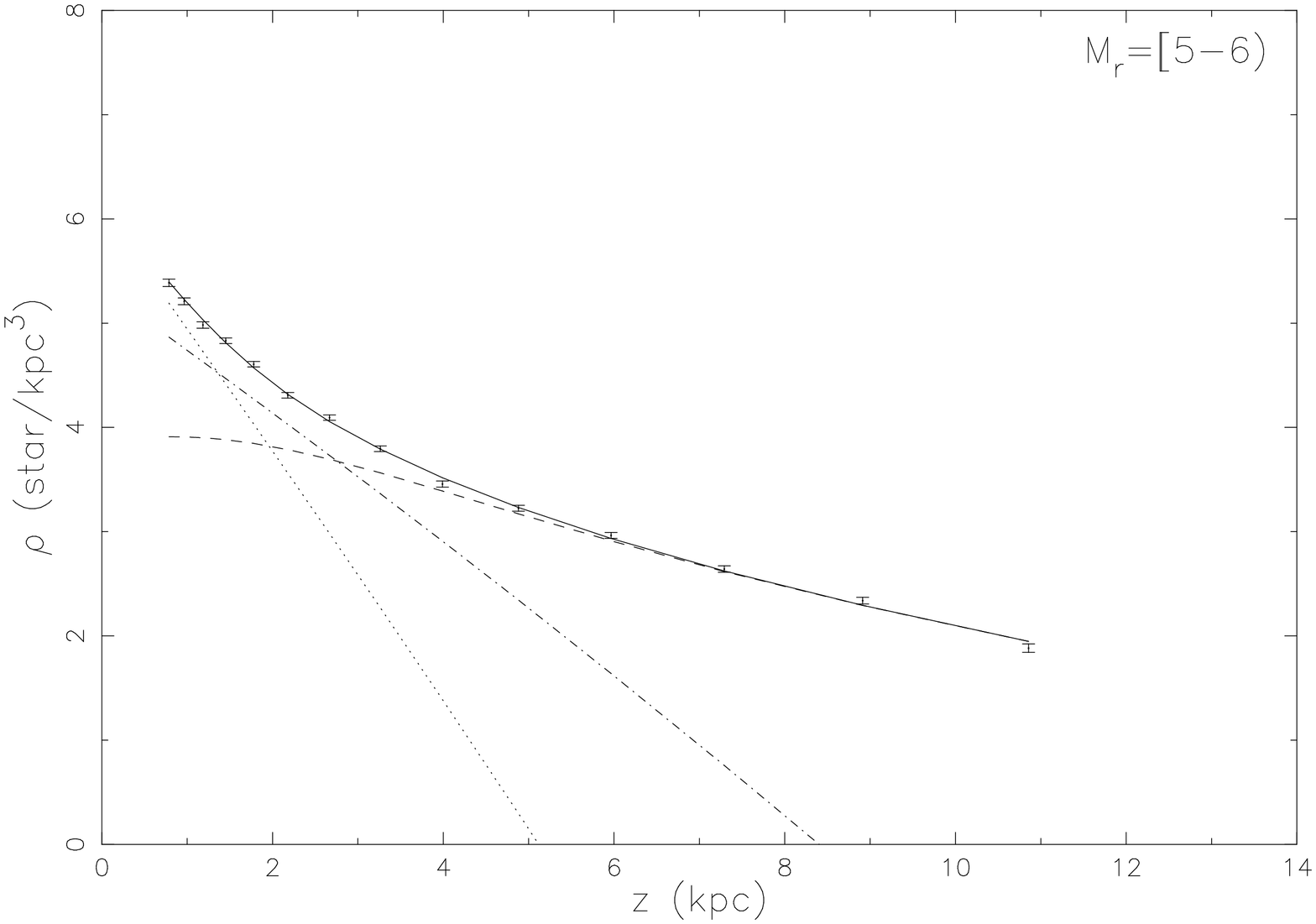}
  \end{minipage}\\[20pt]
  \begin{minipage}[b]{0.45\linewidth}
    \centering
    \includegraphics[angle=-90,width=0.8\textwidth]{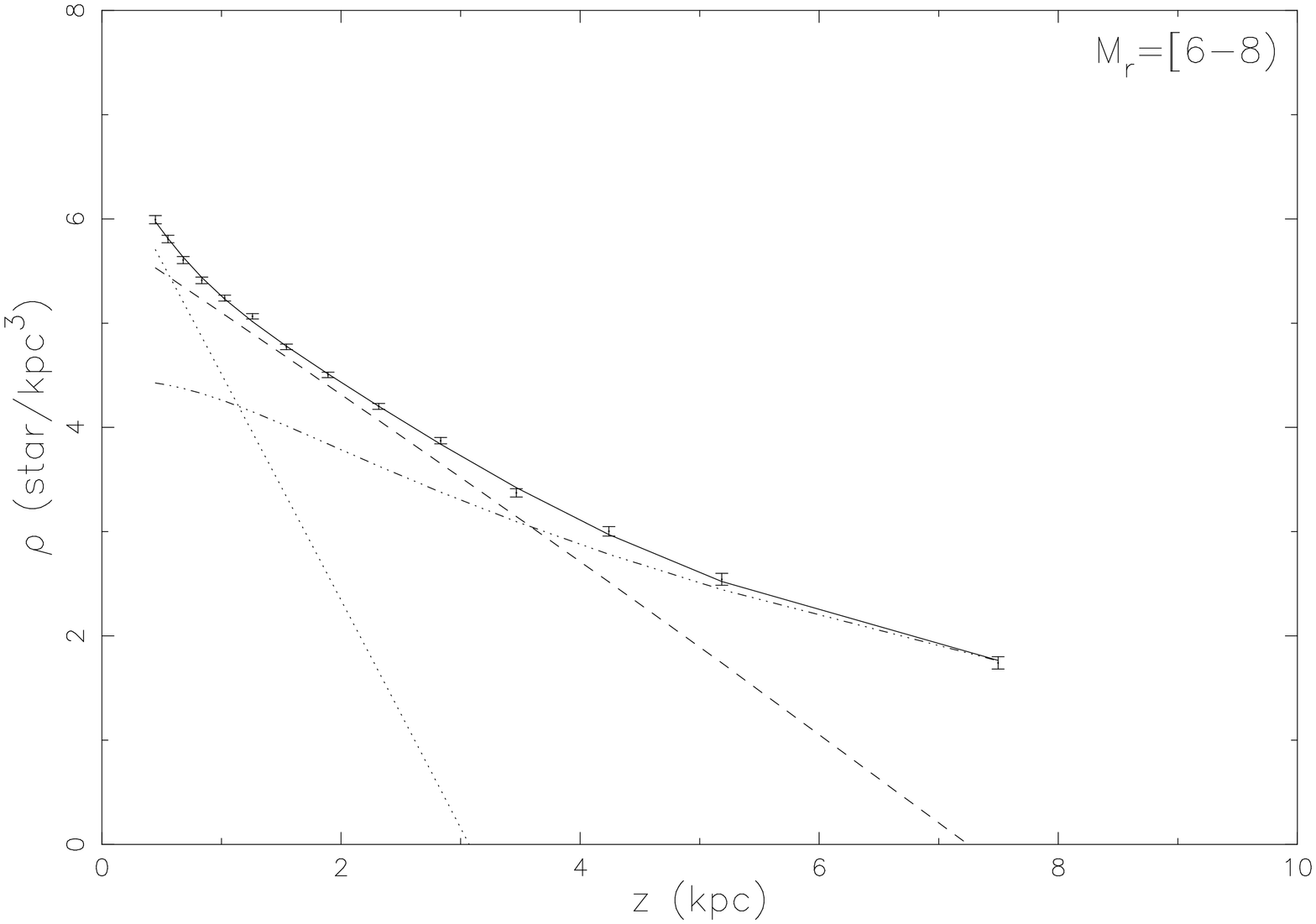}
  \end{minipage}%
  \begin{minipage}[b]{0.45\textwidth}
    \centering
    \includegraphics[angle=-90,width=0.8\textwidth]{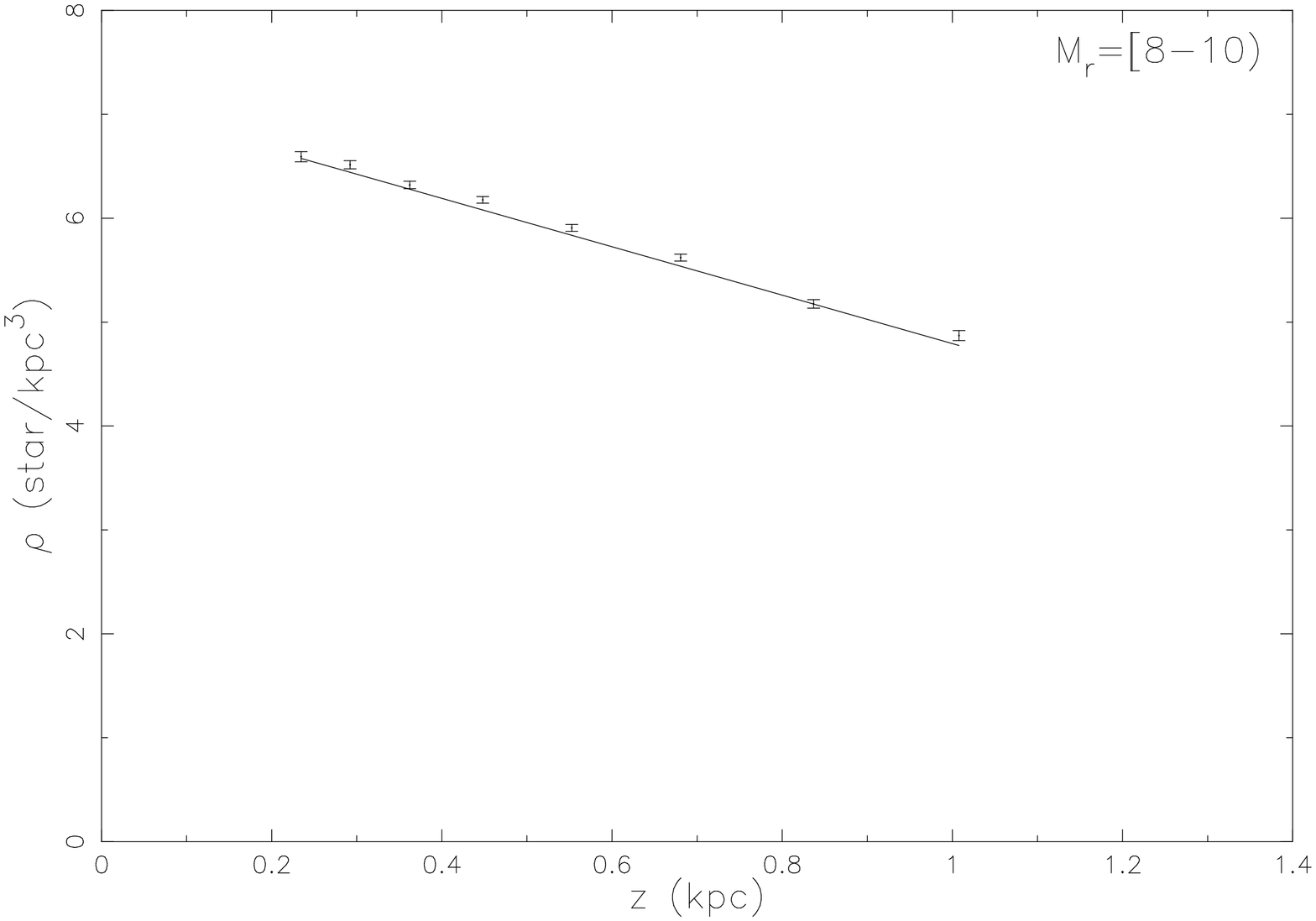}
  \end{minipage}\\[20pt]
  \begin{minipage}[b]{0.45\textwidth}
    \centering
    \includegraphics[angle=-90,width=0.8\textwidth]{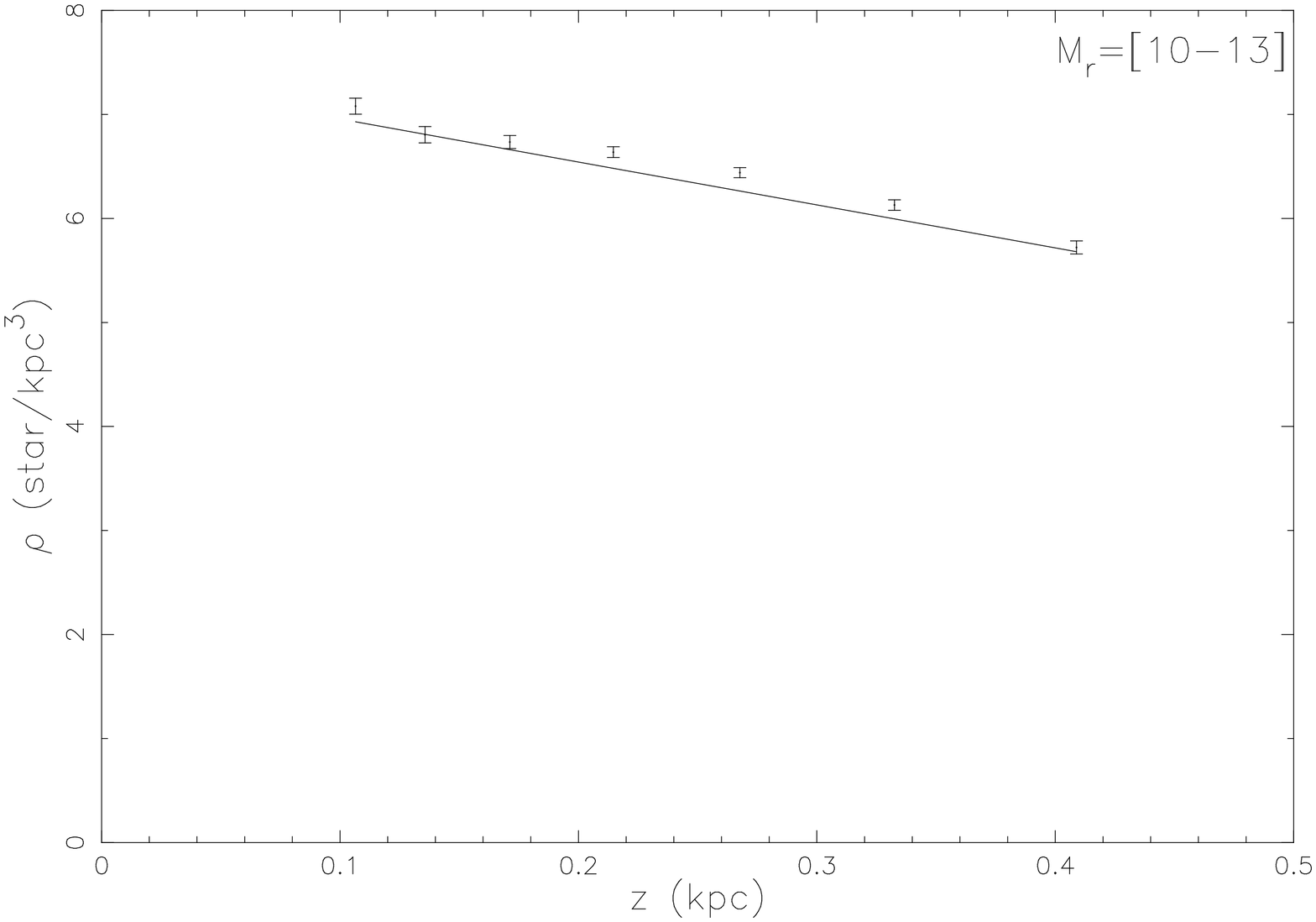}
  \end{minipage}%
  \begin{minipage}[b]{0.45\textwidth}
    \centering
    \includegraphics[angle=-90,width=0.8\textwidth]{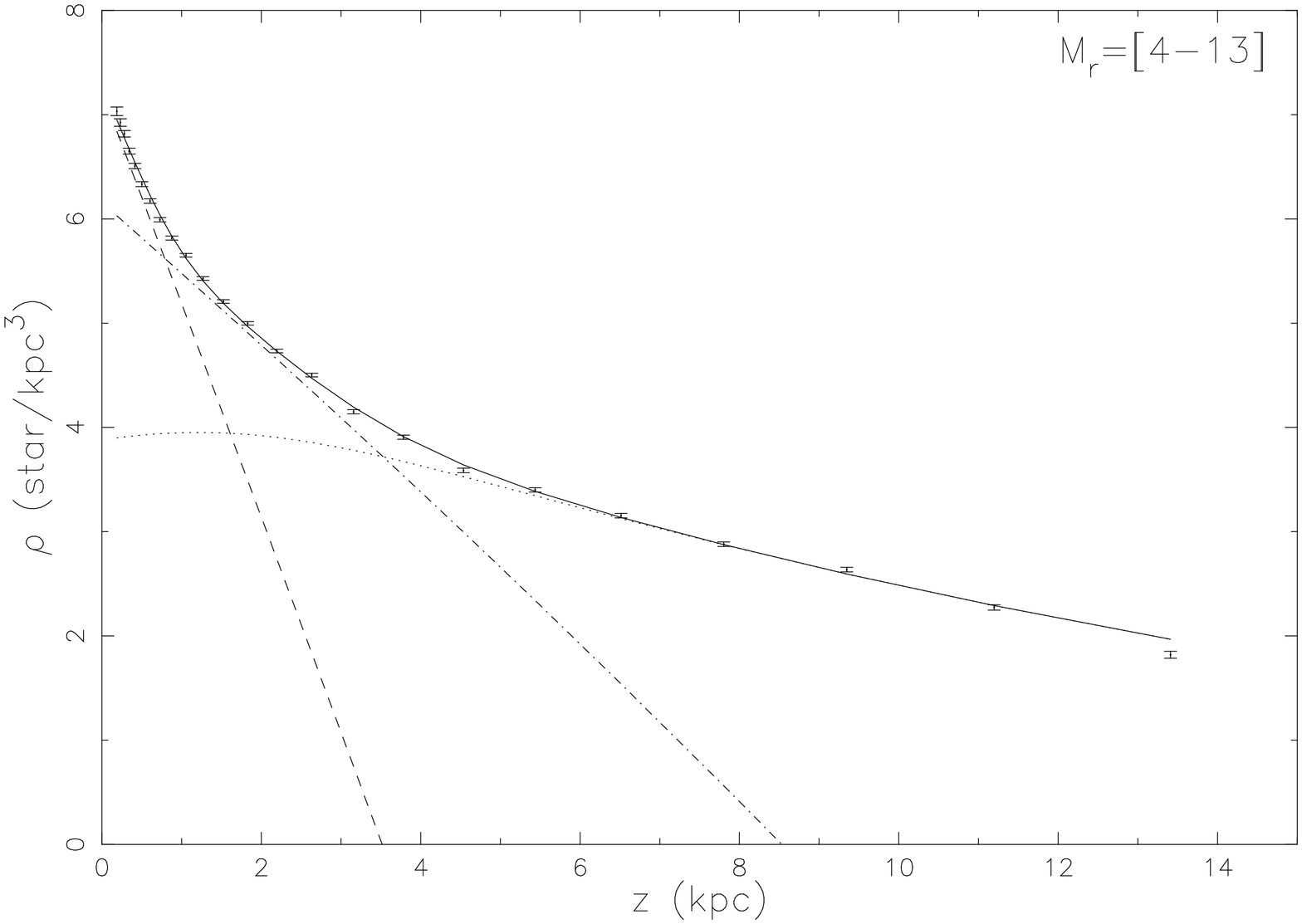}
  \end{minipage}%
\vspace{+0.7cm}
\caption{Observed (dots with error bars) and evaluated (solid line) space density functions
combined for the considered population components which corresponding to
these listed in Table~\ref{tab2}.
 } \label{fig_plot}
\end{figure*}

Figure~\ref{fig_plot} shows the observed and evaluated space density functions
combined for the considered components which corresponding to
these listed in Table~\ref{tab1}.In this figure,
the corresponding absolute magnitude intervals are $4 \leq M_r < 5$,
$5 \leq M_r < 6$, $6 \leq M_r < 8$, $8 \leq M_r < 10$,
$10 \leq M_r \leq 13$, $4 \leq M_r \leq 13$.

\begin{table*}
\begin{center}
  \begin{tabular}{ccccccccc} \hline
{\boldmath{$M_r$}} & \boldmath{$n_1(R_\circleddot,Z_\circleddot)$} & \boldmath{$h_1$} &
\boldmath{$n_2(R_\circleddot,Z_\circleddot)$} & \boldmath{$h_2$} &
\boldmath{$n_2/n_1$} & \boldmath{$n_3(R_\circleddot,Z_\circleddot)$} &
\boldmath{$\kappa$} & \boldmath{$n_3/n_1$} \\
$(mag)$ & $(star/kpc^3)$ & $(pc)$ & $(star/kpc^3)$ &
$(pc)$ & $(percent)$ & $(star/kpc^3)$ & & $(percent)$ \\ \hline
$[4-5)$ & & & $(4.35{+0.48 \atop -0.47})E(5)$ & $575{+20 \atop -20}$
& & $(1.54{+0.14 \atop -0.15})E(3)$ & $0.84{+0.020 \atop -0.040}$ &  \\ \hline
$[5-6)$ & $(1.25{+0.05 \atop -0.05})E(6)$ & $350{+10 \atop -10}$ &
$(2.06{+0.32 \atop -0.31})E(5)$ & $675{+30 \atop -35}$ & $16.5{+2.50 \atop -2.50}$ &
$(7.95{+0.70 \atop -0.75})E(3)$ & $0.45{+0.025 \atop -0.025}$ & $0.64{+0.06 \atop -0.06}$ \\ \hline
$[6-8)$ & $(4.63{+0.25 \atop -0.28})E(6)$ & $194{+12 \atop -12}$ &
$(7.17{+0.70 \atop -0.69})E(5)$ & $525{+15 \atop -15}$ & $15.5{+1.50 \atop -1.50}$
&$(3.01{+0.65 \atop -0.60})E(4)$ & $0.20{+0.025 \atop -0.025}$ & $0.65{+0.14 \atop -0.13}$ \\ \hline
$[8-10)$ & $(1.31{+0.10 \atop -0.09})E(7)$ & $180{+5 \atop -4}$ \\ \hline
$[10-13]$ & $(2.32{+0.27 \atop -0.27})E(7)$ & $103{+5 \atop -5}$ \\ \hline
$[4-13]$ & $(1.66{+0.08 \atop -0.07})E(7)$ & $205{+7 \atop -7}$ &
$(1.37{+0.13 \atop -0.12})E(6)$ & $595{+15 \atop -15}$ & $8.25{+0.75 \atop -0.75}$ &
$(8.23{+1.65 \atop -1.64})E(3)$ & $0.575{+0.025 \atop -0.025}$ & $0.05{+0.01 \atop -0.01}$ \\ \hline
 \end{tabular}
\end{center}
\caption{The Galactic model parameters obtained in this study. $n_i$ and $h_i$ are local density and
scaleheight for the thin and thick disks, respectively.
$\kappa$ is axial ratio of the halo. Suffix 1,2,3,
 denote thin disk, thick disk, halo,respectively. $n_2/n_1$ and $n_3/n_1$ are
the normalization density for thick disk and halo, respectively.
}
\label{tab2}
\end{table*}

\begin{table}
\begin{center}
\begin{tabular}{cccc} \hline
 $M_r$ & $h_2$ & $n_2$ & $n_2/n_1$ \\
($mag$) &  ($pc$) & ($star/kpc^3$) & ($percent$) \\ \hline
[5-6) & 675 (900)  & 2.06E(5) (1.44E(5))  & 16.5 (11.5) \\ \hline
[6-8) & 525 (575)  & 7.17E(5) (6.95E(5))  &15.5 (15.0) \\ \hline
[4-13] & 595 (697)  & 1.37E(6) (1.04E(6))  &8.25 (6.25) \\ \hline
\end{tabular}
\end{center}
\caption{Parameters of thick disk. The values with parenthesis are
obtained by disk-fitting, and those without parenthesis are obtained
by halo-fitting.} \label{tab3}
\end{table}

\section{SUMMARY AND DISSCUSION}

We estimate the Galactic model parameters using $\chi^2$ method in absolute
magnitude intervals: $4 \leq M_r < 5$, $5 \leq M_r < 6$, $6 \leq M_r < 8$,
$8 \leq M_r < 10$, $10 \leq M_r \leq 13$, with a unique density law for each
population individually of 7.07 $deg^2$ field locate at
$50 ^\circ \leq l \leq 55^\circ$, $-46 ^\circ \leq b\leq -44^\circ $, to explore
their possible variation with absolute magnitude from SDSS and SCUSS photometry.
We also estimate the parameters in absolute magnitude interval $4 \leq M_r \leq 13$ in
order to compare with other researchers' works. Our results show the parameters
are absolute magnitude dependent (see Table~\ref{tab2}). This is a possible
reason to explain why different researchers obtained different parameters.

\subsection{Thin disk}

The local density of the thin disk increases and the scaleheight decreases,
 with the sample stars get fainter. The local density
$n_1(R_\circleddot,z_\circleddot)$ varies from $1.25\times10^6$ $star/kpc^3$ in absolute magnitude interval
$5 \leq M_r < 6$ to $2.32\times10^7$ $star/kpc^3$ in absolute magnitude interval $10 \leq M_r \leq 13$,
and the scaleheight changes from 350 $pc$ to 103 $pc$ in the corresponding
absolute magnitude interval. The results reveal a phenomenon for thin disk:
the intrinsic faint MS stars tend to stay closer to the Galactic plane,
 and have  larger local density. In order to compare with other researchers'
results, we derive the scaleheight of the thin disk in absolute magnitude
interval $4 \leq M_r \leq 13$, that is 205 $pc$, which is a little
 smaller than most previous work (see Table~\ref{table_ref}).
 As these parameters depend on absolute magnitude, we can conclude that
 the small scaleheight of thin disk in absolute magnitude interval $4 \leq M_r \leq 13$ is
caused by the contribution of the faint MS stars in our sample.

It is obvious that different types of MS stars could have experienced
different dynamic process, thus could have different specific distributions.
In other words, the specific distribution of MS stars could dependent on
absolute magnitude.
The quantitative calculation to explain this phenomenon may
need the chemical, kinematic and dynamic information,
the star formation history and star formation rate may also be included.
Here we just give a possible qualitative explanation:
bright MS stars have large mass and short
lifetime which lead the local density to be relatively smaller than that of the faint one ,
thus the gravitational interaction time they experienced are short,
this may make the bright MS stars stay a little further in mean than the
fainter one, so the scaleheight $h_1$ of the bright MS stars would
be larger.

\subsection{Thick disk}

The final results of the thick disk suffer from the degeneracy between
the two halo parameters: $\kappa$ and $n_3$, see Figure~\ref{fig_halocontour}.
 It raise the uncertainty of the relationship between thick disk parameters
and absolute magnitude.
\textbf{As our local density of thin disk $n_1$
 changes with absolute magnitude,  it is significative to compare
 the value of $n_2$ with other works rather than $n_2/n_1$.
The values of $h_2$ we obtained are a little smaller than previous works,
 thus the value of $n_2$ (normalized at $(R_\circleddot,z_\circleddot)$) should be a little larger.
As shown in Table~\ref{tab2}, $h_2$ roughly fall into the range of  $520  < h_2 (pc) < 680$, simultaneously,
$n_2$ roughly fall into the range of
$2.0\times10^5 < n_2 (star/kpc^3) < 1.4\times10^6$.
The results show the parameters of thick disk are non-monotonic
change with absolute magnitude, which may imply the origin of
the thick disk is complicated.}

\subsection{Halo}

As shown in Table~\ref{tab2}, the halo parameters $\kappa$ and $n_3$ are monotonic
 change with absolute magnitude. The axial ratio $\kappa$ varies from 0.84 in bright
interval to 0.20 in faint interval which indicates a flattened inner halo and
a more spherical halo. The trend of $n_3$ shows intrinsic faint MS stars have large
local density.

In general, the values of $\chi^2$ for halo-fitting are larger than disk-fitting
in all intervals. Figure~\ref{fig_plot} can reflect the large deviation
of halo-fitting.  The deviation  may come from  the influence of the
uncompleted  data, especially at large distance.

As seen in Figure~\ref{fig_halocontour},
the degeneracy exists between the halo parameters $n_3$ and $\kappa$.
 So the confidence of the halo parameters
monotonic change with absolute magnitude will be influenced. We do not know
what cause the halo parameters degenerate, but the most likely reason is
 the unsatisfactory of the halo component fitting: whether the observation
is not good enough (e.g. the incomplete data in halo component) or the halo model is not
precisely describe the Galaxy.

In summary, we fit the observations with three components Galactic model
to estimate the Galactic structure parameters to explore their possible variations
with absolute magnitude. Our results show the parameters of the thin disk and halo
are monotonic change with absolute magnitude, but the thick disk is
non-monotonic. The explanation of this phenomenon need more work.
Also, we still need a new method to estimate the Galactic structure parameters
in order to break the degeneracy.
 As greatly improve on data collection and more and more researchers work on these field
  in recent years, we believe these questions
 can be solved in the near future. We still have a long way to know our galaxy well.

\section*{ACKNOWLEDGMENTS}

We especially thank the referee, S. Bilir,  for his insightful comments and suggestions
which have improved the paper significantly.
This work was supported by joint fund of Astronomy of the
the National Nature Science Foundation of
China and the Chinese Academy of Science, under Grants U1231113.
This work was also supported by the GUCAS president fund
and the Chinese National Natural Science Foundation grant No. 11373033.
This work has been supported by the Chinese National Natural Science Foundation
through grants 11373035, and by the National Basic Research Program of China (973 Program), No. 2014CB845702.

We would like to thank all those who participated in observations
and data reduction of SCUSS
for their hard work and kind cooperation.
The SCUSS is funded by the Main Direction Program of Knowledge Innovation of Chinese Academy of Sciences (No. KJCX2-EW-T06). It is also an international cooperative project between National Astronomical Observatories, Chinese Academy of Sciences and Steward Observatory, University of Arizona, USA. Technical supports and observational assistances of the Bok telescope are provided by Steward Observatory. The project is managed by the National Astronomical Observatory of China and Shanghai Astronomical Observatory.

Funding for SDSS-III has been provided by the Alfred P. Sloan Foundation, the Participating Institutions, the National Science Foundation, and the U.S. Department of Energy Office of Science. The SDSS-III web site is \emph{http://www.sdss3.org/}.

SDSS-III is managed by the Astrophysical Research Consortium for the Participating Institutions of the SDSS-III Collaboration including the University of Arizona, the Brazilian Participation Group, Brookhaven National Laboratory, Carnegie Mellon University, University of Florida, the French Participation Group, the German Participation Group, Harvard University, the Instituto de Astrofisica de Canarias, the Michigan State/Notre Dame/JINA Participation Group, Johns Hopkins University, Lawrence Berkeley National Laboratory, Max Planck Institute for Astrophysics, Max Planck Institute for Extraterrestrial Physics, New Mexico State University, New York University, Ohio State University, Pennsylvania State University, University of Portsmouth, Princeton University, the Spanish Participation Group, University of Tokyo, University of Utah, Vanderbilt University, University of Virginia, University of Washington, and Yale University.


\end{document}